\documentclass[letter,twocolumn]{jpsj3}

\usepackage{txfonts}

\usepackage{amsmath,amssymb}
\usepackage{empheq}
\usepackage{graphicx}
\usepackage{bm}
\usepackage{color}

\usepackage{ulem}

\usepackage{dcolumn}

\newcommand{\va}{ \bm{a} }

\newcommand{\vd}{ \bm{d} }
\newcommand{\vdd}[1]{ \bm{d}_{#1} }
\newcommand{\vk}{ \bm{k} }

\newcommand{\vR}{ \bm{R} }
\newcommand{\vS}{ \bm{S} }
\newcommand{\vtau}{ \bm{\tau} }

\newcommand{\PLUS}[2]{{\mbox{$#1$$+$$#2$}}}

\title{Nematicity Liquid in a Trimerized-Kagome Antiferromagnet}

\author{Ichiro Tanaka and Hirokazu Tsunetsugu}
\inst{
The Institute for Solid State Physics, 
The University of Tokyo, 
Kashiwanoha 5-1-5, Chiba 277-8581, Japan 
} 

\abst{
We theoretically study low-temperature properties 
of the antiferromagnetic Heisenberg model with half-integer spin $S$
on the kagome lattice with large trimerization.  
We have derived a low-energy effective model for general $S$ in terms of 
spin and nematicity operators in triangular units, and 
studied their low-temperature correlations 
for $S$=$3/2$ by classical Monte Carlo simulations.  
The previous study for the $S$=$1/2$ case 
[M.~Ferrero \textit{et al.}, 
Phys.~Rev.~B \textbf{68}, 214431 (2003) ] 
reported a spin liquid state at low temperatures driven by 
a glassy behavior of isolated dimers and trimers of nematicities. 
The results for $S$=$3/2$ show that nematicity  
dimers and trimers are connected by weak links to form 
a defective planar network. 
At very low temperatures, nematicities show glassy 
slow dynamics, and cluster spins continue to fluctuate 
in a nonperiodically frozen background of nematicities.   
The characteristic time of glassy dynamics scales 
with temperature following a power law.  
}

\begin{document}
\maketitle

Heisenberg antiferromagnets on the kagome lattice 
is the representative frustrated spin system, 
and 
the possibility of quantum spin liquid ground state has 
been intensively studied for many decades 
both theoretically and experimentally, 
but its conclusion remains under debate.\cite{kagome1,kagome2}  
For its better understanding, it is useful to 
study a related system with frustration more tamed, 
and we study in this letter the model on 
a trimerized kagome lattice, which is also called 
breathing kagome.  
This system is also interesting in itself 
due to a few instances of its realization.  
The vanadate 
(NH$_{4}$)$_{2}$[C$_{7}$H$_{14}$N][V$_{7}$O$_{6}$F$_{18}$] (DQVOF) 
has a trimerized kagome network of V$^{4+}$ ions with spin $S$=1/2, 
and a spin liquid behavior is observed.\cite{Clark,Orain}
There is also a proposal of an optical lattice 
with trimerized kagome geometry.\cite{Santos}
Recently, Ishii \textit{et al.} synthesized 
Li$_{2}$Cr$_{3}$SbO$_{8}$, in which  
Cr$^{3+}$ ions with $S$=3/2 form a trimerized kagome network,  
and observed a $\sqrt{3}$$\times$$\sqrt{3}$ 
magnetic order and plateaus in its magnetization curve.\cite{Iida,Ishii}
A few theoretical studies discussed that its ground state 
is a spin liquid.\cite{Ferrero,Santos,Schaffer,Iqbal}

One of the authors studied the Heisenberg antiferromagnets 
on a breathing pyrochlore lattice.\cite{Tsunetsugu01,Tsunetsugu17} 
An important difference from kagome lattices  
lies in the parity of the number of 
spins in unit cell.  
Since the pyrochlore lattice has a tetrahedron unit, the parity is even.
Its ground states have degeneracy $2S$$+$$1$,  
and they are all spin-singlet irrespective of spin $S$.  
By contrast, the parity is odd for the kagome lattice.
For integer $S$, each triangular unit has a unique ground state.  
Therefore the ground state of the whole system 
is a valence bond crystal, and no symmetry is broken.  
The case of half-integer $S$ is interesting and 
the ground state in each unit 
has degeneracy 4$=$2$\times$2, i.e. two spin doublets.\cite{Mila98}  
This additional nonmagnetic 2-fold degrees of freedom is related to 
the point group symmetry and let us call it \textit{nematicity}, which 
will be explained more explicitly later.  
The issue to study is to see how these cluster spins and nematicities 
in the triangular units develop correlations at low temperatures.  
Ferrero \textit{et al.} investigated this problem 
for the $S$=1/2 case.\cite{Ferrero}
Performing classical Monte Carlo (MC) simulations for its low-energy 
Hamiltonian, 
they observed a glassy behavior of nematicity at very low temperatures.
Note that they used the term \textit{``chirality''}, 
which is synonymous with our nematicity.  
In this letter, we will study how the low temperature 
physics changes for higher $S$.  
We will construct an effective Hamiltonian for general half integer $S$ 
and perform classical MC simulations.  

\begin{figure}[bt]
\begin{center}
\includegraphics[width=7.0cm,bb=0 0 567 259]{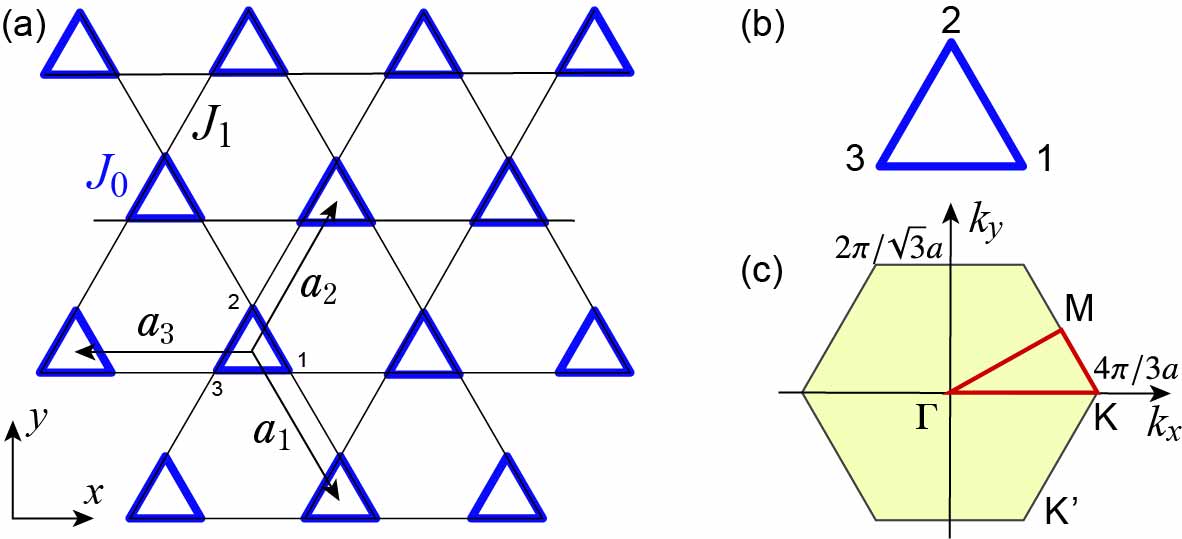}
\end{center}
\caption{(Color online)
(a) Trimerized kagome lattice, aka breathing kagome.  
With lattice constant $a$, 
$\va _1$=$a(1/2,-\!\sqrt{3}/2)$, 
$\va _2$=$a(1/2,\!\sqrt{3}/2)$, and 
$\va _3 $=$-$$\va _1$$-$$\va _2$.
(b) Sublattice indices in unit cell. 
(c) Brillouin zone.  
}
\label{fig1}
\end{figure}

The model to study is 
the trimerized-kagome antiferromagnetic Heisenberg (TKAFH) Hamiltonian 
with half-integer spin $S$ 
\begin{equation}
 \tilde{H} = 
J_0 \sum_{\vR} \sum_{1 \le i < j \le 3} \!\!\!\!
\vS _i (\vR ) \cdot \vS _j (\vR )
+ J_1 \sum_{\vR} \sum_{j=1}^3 
\vS _j (\vR ) \cdot \vS _{j+1} (\mbox{$\vR$+$\va _j$} ) , 
\label{eq:ham0}
\end{equation}
where $J_0$ and $J_1$ denote the nearest-neighbor exchanges 
in triangles pointing up and down respectively
as shown in Fig.~\ref{fig1}(a).  
We will consider the case of strong trimerization 
$0$$<$$J_1$$\ll$$J_0$ in this letter.  
In the following, let us call up-pointing triangles 
\textit{clusters}.  
The sum on $\vR$ is taken over clusters, 
which constitute a triangular lattice with the lattice 
vectors $\va _1$ and $\va _2$.  
$\vS_j (\vR )$ denotes the spin on the sublattice $j$ 
shown in Fig.~\ref{fig1}(b) 
with the convention $\vS_{j + 3} (\vR )$=$\vS_j (\vR )$, 
and define the cluster spin by 
$\vS (\vR )$$\equiv$$\vS_1 (\vR )$+$\vS_2 (\vR )$+$\vS_3 (\vR )$. 

In the decoupled limit $J$=0, the ground state is 
a direct product of cluster ground states.  
Each cluster has 4-fold degenerate ground states 
$|\Phi^{p}(M) \rangle$, ($p$=$\pm$, $M$=$\pm 1/2$), 
for any half-integer spin $S$.  
They are spin doublet with $S^z$=$M$, and 
eigenstates of $(\mbox{$\vS _1$$+$$\vS _3$})^2$ with 
eigenvalue $(\mbox{$S$+$p/2$} )(\mbox{$S$+$p/2$+1})$. 
This doublet $|\Phi^{\pm} \rangle$ constitutes a basis of 
E-irreducible representation of the C$_{3v}$ point group 
of the triangular unit, 
and therefore describes breaking of the lattice symmetry.  
For later use, it is convenient to use the chiral basis: 
$|\Phi_{L,\, R}(M) \rangle$=
$2^{-1/2} \bigl[ |\Phi^{+}(M) \rangle$%
$\,\pm\,$$i \,|\Phi^{-}(M) \rangle \bigr]$.
After lengthy calculations of Clebsh-Gordan coefficients 
and 6-$j$ symbols,\cite{Messiah} 
we can represent the sublattice spin operators 
in terms of this chiral basis 
within the ground state manifold as 
\begin{equation}
{\vS}_j (\vR ) \rightarrow 
\zeta \, \vS (\vR ) \, 
\bigl[ \mbox{$\gamma$$+$$\vdd{j}$$\cdot$$\vtau (\vR )$} \bigr] , 
\ \ 
\vdd{j} = ( \cos \omega_j , \sin \omega_j ) , 
\label{eq:Stau}
\end{equation}
where 
$3\omega_3$=$-3\omega_1$=$\omega_2$=$\pi$,
$\zeta$=$(\mbox{$2S$$+$$1$})/3$ and 
$\gamma$=$1/(\mbox{$2S$$+$$1$})$$<$$1$.  
Note that while $\vS _j (\vR )$ are spin-$S$ operators, 
the cluster ones $\vS (\vR )$ are spin-1/2.  
$\vtau$=$(\tau_1 , \tau_2 )$ are Pauli matrices 
operating in the chiral space $(\Phi_L, \Phi_R)$ 
at each unit $\vR$. 
These operators constitute a basis set of E-irreducible 
representation of the triangular point group, and 
describe breaking of this symmetry. 
Therefore, we will call them \textit{nematicity} in this letter.  
Note that the chirality is represented by the other operator
$\tau_3$=$i\tau_2 \tau_1$.\cite{Tsunetsugu01} 

The low-energy effective Hamiltonian of this system is 
derived by projecting $\tilde{H}$ into the macroscopically degenerate 
ground states of the $J_0$ term.  
Dropping a constant term, 
it is the $J_1$ term represented with the projected 
operators (\ref{eq:Stau}) 
\begin{align}
&H = 
J \sum_{\vR} \sum_{j=1}^3 \ 
\vS (\vR ) \cdot \vS (\mbox{$\vR$+$\va _j$} ) \  
\Gamma (\vR , \mbox{$\vR$+$\va _j$} ) , 
\ \ \ (\mbox{$J$=$\zeta^2$$J_1$}) 
\label{eq:ham}
\\[-4pt]
&\Gamma (\vR , \mbox{$\vR$+$\va _j$} ) 
\equiv 
 \bigl[ \mbox{$\gamma$$+$$\vdd{j}$$\,\cdot\,$$\vtau (\vR )$} \bigr] 
 \bigl[ \mbox{$\gamma$$+$$\vdd{j+1}$%
$\cdot\,$$\vtau ($}\mbox{$\vR$+$\va _j$}) \bigr] . 
\label{eq:Gamma}
\end{align}
This effective Hamiltonian was previously obtained for 
the $S$=$1/2$ case by Mila.\cite{Mila98}
It is important that 
the Hamiltonian does not change its form for higher $S$'s,\cite{Mambrini}  
but the value of $\gamma$ varies 
in addition to the renormalization of the energy unit $J$.
We will show later that the change of $\gamma$ qualitatively modifies 
the correlations of $\vtau$'s from the $S$=1/2 case.  
This approach is based on the first order perturbation,  
and justified as far as the renormalized energy scale $J$ is 
small enough compared with the energy gap $J_0$ 
in the decoupling limit.  

\begin{figure}[bt]
\begin{center}
\includegraphics[width=8.5cm,bb=0 0 1419 495]{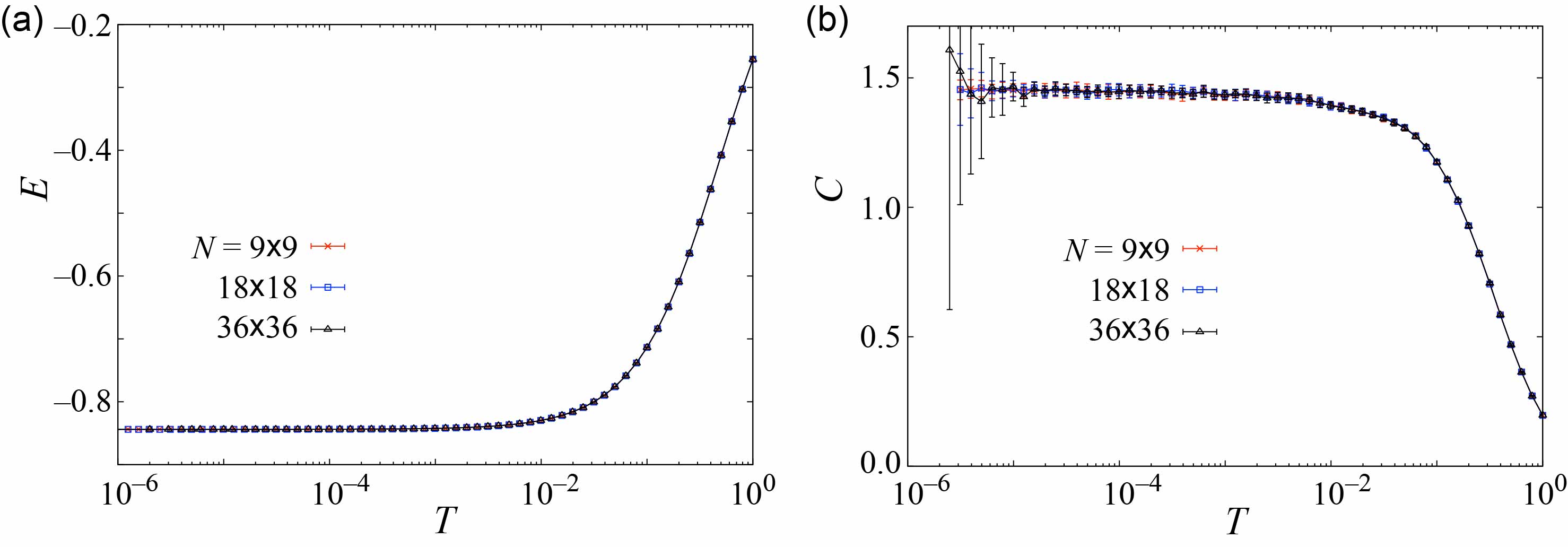}
\end{center}
\caption{(Color online)
(a) Internal energy and (b) specific heat 
of the effective model $H$ for $S$=3/2.  
Values are normalized per site, \textit{i.e.} 
three spins in $\tilde{H}$.  
Error bars are estimated by the bootstrap method 
with 95\% percentile.  
}
\label{fig2}
\end{figure}

In order to examine how their correlations evolve with 
lowering temperature $T$, 
we have investigated the effective model $H$ by classical 
MC simulations.  
Namely, spins $\{ \vS (\vR ) \}$ and nematicities $\{ \vtau (\vR ) \}$ 
are treated as \textit{classical} unit vectors 
with 3 and 2 components, respectively.
In the following, we use $J$ as the units of 
energy and set $J$=1.  
The previous study for $S$=$1/2$ used a similar approximation 
but treated $\{ \vtau (\vR ) \}$ as 3-component 
classical vectors.\cite{Ferrero}
However, this difference does not change results qualitatively, 
since nematicities hardly point to the third direction 
at low $T$.  
Simulated systems are rhombus parts of the triangular 
lattice with edges parallel to $\va _1$ and $\va _2$, 
and periodic boundary conditions are imposed.  
The number of sites $N$ is up to 36$\times$36, 
and we use the standard Metropolis algorithm with local updates.  
Physical quantities are measured typically for $10^6$
Monte Carlo steps (MCS) following $10^6$ MCS of annealing 
for each temperature, 
and they are averaged over 128 different ensembles.  

We first calculated thermodynamic quantities 
to check a possibility of phase transition. 
Since the spin part of $H$ is isotropic, 
we do not expect a finite-temperature phase transition in the spin space.  
However, since the nematicity interactions are anisotropic, 
a possibility of phase transition remains.  
Figures \ref{fig2}(a) and (b) plot 
internal energy $E(T)$ and specific heat $C(T)$, respectively, 
of the effective Hamiltonian (\ref{eq:ham}).  
The $C(T)$ curve gives no indication of diverging peak 
down to the very low temperature 
$T$$\sim$$10^{-6}$.  
The internal energy provides no indication of jump either. 
These results imply that no phase transition occurs, 
neither first-order or continuous one, and 
the lowest-temperature state remains disordered.  
In contrast to the results in Ref.~8,  
$C(T)$ exhibits no peak, but 
it is common that the nearly flat part has a value $\sim$1.5, 
although their $\vtau$ has 3 components. 
We will discuss this point later.  

\begin{figure}[bt]
\begin{center}
\includegraphics[width=8.0cm,bb=0 0 1407 507]{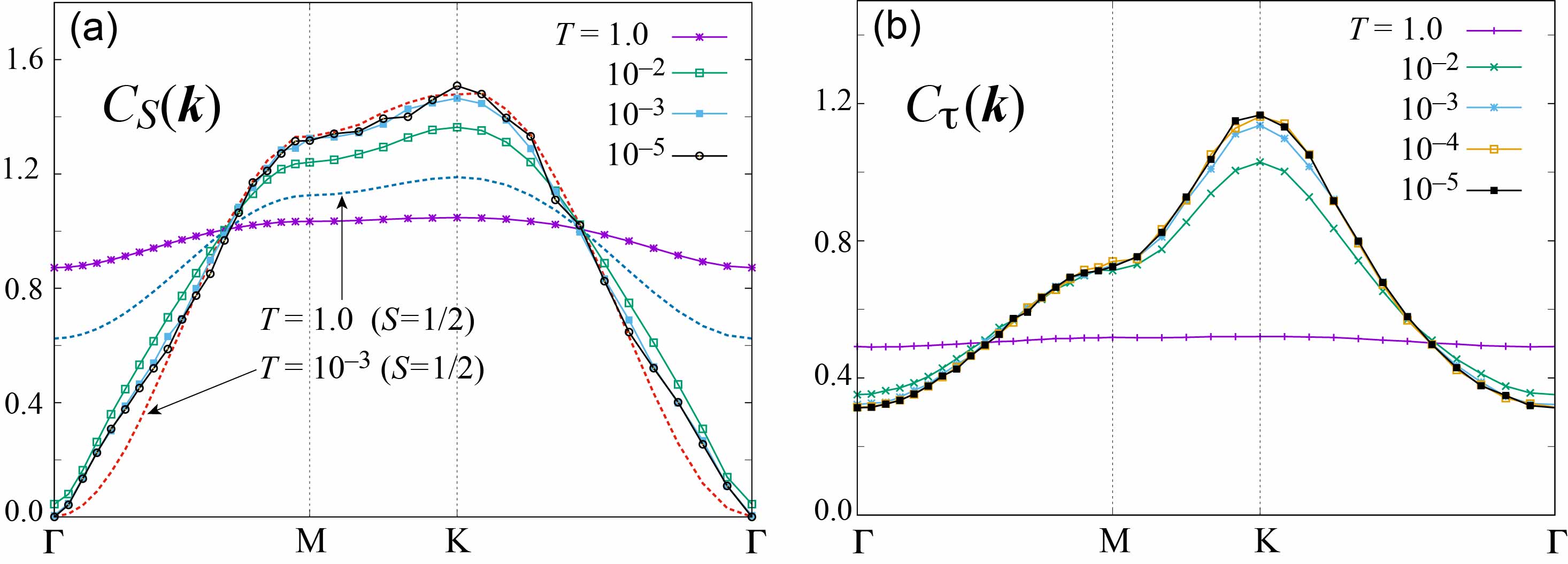}
\end{center}
\caption{(Color online)
(a) Spin correlations $C_S (\vk ) $ and  
(b) nematicity correlations $C_\tau (\vk ) $ 
of the effective model $H$ for $S$=$3/2$ and $N$=36$\times$36.  
For comparison, $C_S (\vk ) $ for $S$=$1/2$ is also calculated 
and shown in (a). 
}
\label{fig3}
\end{figure}

We next investigate correlations of spins and nematicities.  
Two-dimensional systems with continuous internal 
symmetry often undergo a phase transition at $T$=0, 
and in such a case a related correlation function 
shows a divergent temperature dependence.\cite{Kawamura1984}  
Define spin and nematicity correlation functions as 
$
 C^{\mu \nu }_O (\vk ) $$\equiv$%
$ N^{-1} \sum_{\vR , \vR '} 
\langle O_{\mu} (\vR ) O_{\nu} (\vR ' )  \rangle$
$ e^{-i \vk \cdot (\vR - \vR ')}$, 
($O$=$S$, $\tau$) 
and let $C_S (\vk )$=$\sum_\mu C^{\mu \mu}_S (\vk)$ and 
$C_\tau (\vk )$ denote the larger eigenvalue of the latter one.  
Note that measured $\langle S_{\mu} (\vR ) S_{\nu} (\vR ') \rangle$ 
is proportional to the identity matrix within statistical errors.  
We have calculated $C_S (\vk )$ for all $\vk$'s, 
and plot the results along the high symmetry axes in Fig.~\ref{fig3}(a).  
It has a broad peak at the Brillouin zone corners (K and K')
at all $T$'s, which indicates 
antiparallel or a 120$^\circ$-type spin correlation between 
nearest neighbor sites.  
The peak does not grow or narrow with lowering temperature.  
This sharply contrasts with the behavior of the growing 
120$^\circ$ structure in the antiferromagnetic Heisenberg 
model on the triangular lattice.\cite{Kawamura1984,Kawamura2010}

\begin{figure}[bt]
\begin{center}
\includegraphics[width=8.0cm,bb=0 0 991 1001]{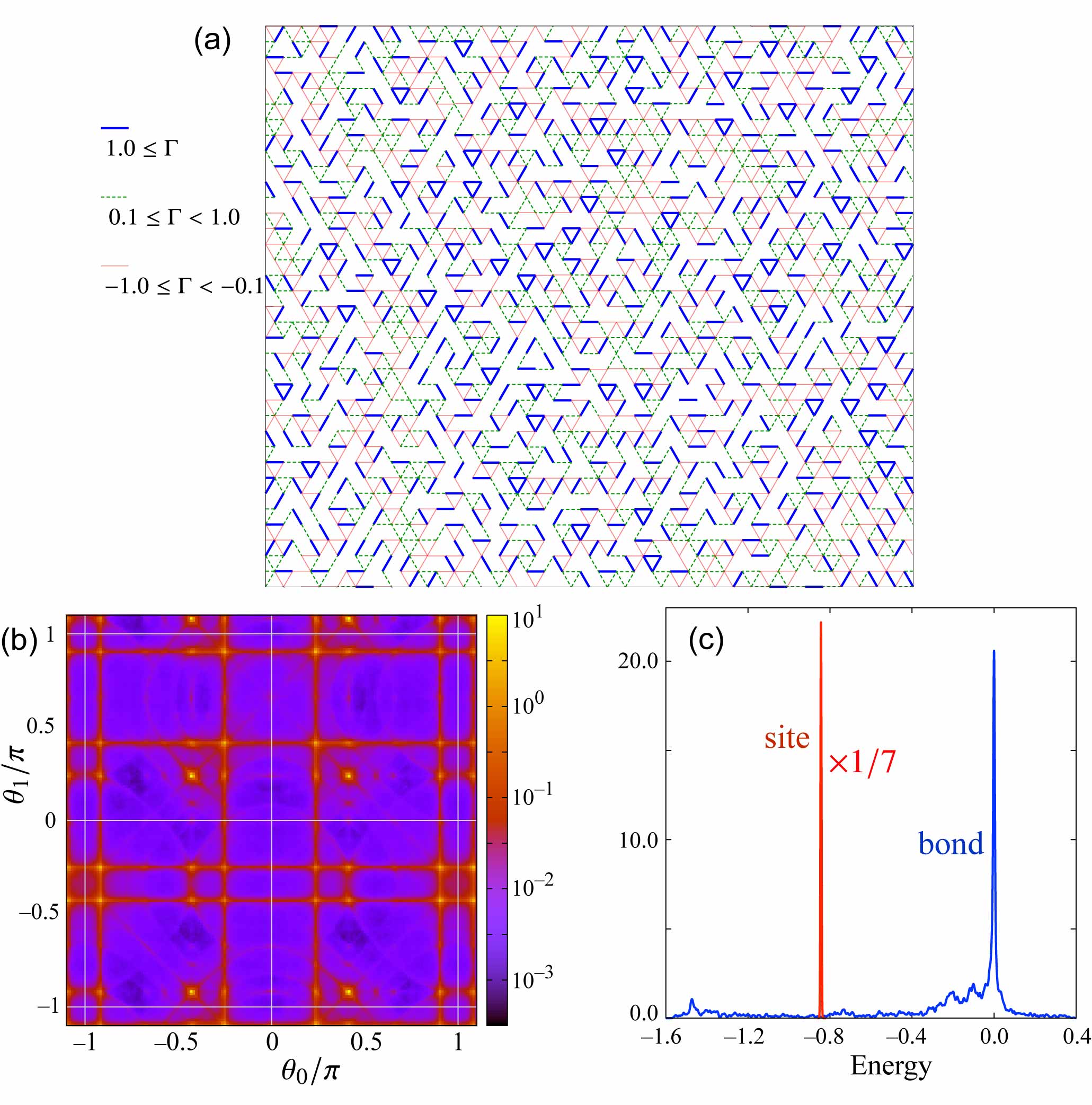}
\end{center}
\caption{(Color online)
(a) Snapshot of nematicity correlations $\Gamma (\vR , \vR ')$ for 
the $S$=$3/2$ case at $T$=$10^{-5}$. 
The system size is $N$=36$\times$36.  
(b) Pair distribution $P(\theta_0,\theta_1)$ 
calculated at $T$=$10^{-5}$ with averaging over 6144 ensembles.  
(c) Distribution function of 
bond and site energies for the snapshot (a).  
Each peak is broadened by a Gaussian with width $0.002$ and $0.005$ 
for site and bond energy, respectively.  
}
\label{fig4}
\end{figure}

Nematicity correlation function $C_\tau (\vk )$ is plotted 
in the same way in Fig.~\ref{fig3} (b).  
Compared to the spin correlations, the peak at K point is more prominent. 
However, it remains broad 
and its width and height barely change at low temperatures.  
These results clearly demonstrate that the effective Hamiltonian 
(\ref{eq:ham}) exhibits no tendency of long-range order in 
not only spin but also nematicity spaces.  
In this sense, the system remains in the liquid phase 
of spin and nematicity.  

The previous study for the $S$=$1/2$ case pointed out that 
this low-temperature state has a unique character
in their local nematicity correlations.\cite{Ferrero}
To quantify this, let us examine for each bond 
$\Gamma (\vR , \mbox{$\vR$+$\va _j$} )$ defined in Eq.~(\ref{eq:Gamma}), 
which is bounded as
$\Gamma_-$$\le$$\Gamma (\vR , \vR ')$$\le$$\Gamma_+$ with 
$\Gamma_-$=$-(\mbox{1$-$$\gamma^2$})$$<0$ and 
$\Gamma_+$=$(\mbox{$1$+$\gamma$})^2$.  
Each site is connected to 6 bonds, 
and MC snapshots for the $S$=$1/2$ case showed 
a unique imbalance in $\Gamma$ 
among these 6 bonds.\cite{Ferrero}
At majority of sites, one and only one bond has a nonvanishing value 
$\approx \Gamma_+$, and 
those bonds were called \textit{dimers}. 
At most of the remaining sites, $\tau$'s form \textit{trimers}, 
which are called triangles in the previous study\cite{Ferrero}. 
Three sites are connected by three bonds with $\Gamma_{+}$ 
to form a triangular unit, 
while $\Gamma$=0 for the remaining 12 bonds.  

For the $S$=$3/2$ case, 
a MC snapshot of nematicity correlations $\Gamma (\vR , \vR ')$ 
is plotted in Fig.~\ref{fig4}(a). 
It exhibits similarities and important differences compared with 
the $S$=$1/2$ case.  
Bonds with large $\Gamma$ value are shown by thick blue line, 
and the most noticeable similarity is that 
most of the sites belong to a dimer or a trimer.  
An important difference is that many of the remaining bonds 
have a sizable $\Gamma$, and let us call them \textit{weak links}.  
Thus, dimers and trimers are not isolated but connected 
by these weak links to form a \textit{defective planar network} 
with modulated bond strengths 
mixed with small disconnected blobs such as 
unglued dimers and hexamers.  

Spins fluctuate on this network with modulated effective 
exchange couplings $\{ \Gamma (\vR , \vR ') \}$,
which are predominantly antiferromagnetic.
Since the network extends over the whole system, 
spins establish long-range correlations, 
and this is manifested by $C_S (\vk )$$\sim$$|\vk |$
around $\Gamma$ point as shown in Fig.~\ref{fig3}(a).  
This nonanalyticity indicates 
a power-law scaling of long-range asymptotics 
in envelope 
$| \langle \mbox{$\vS (\vR )$$\cdot$$\vS (\bm{0})$} \rangle | $%
$\sim$$\Lambda^{1/2}R^{-5/2}$, 
where $\Lambda$ is the wavenumber cutoff.
By contrast, the $S$=1/2 result is smooth, 
$C_S (\vk )$$\sim$$\vk ^2$, 
and this is a consequence of short-range spin correlations 
confined within isolated dimers and trimers. 

Let us investigate the origin of these local nematicity correlations 
at low $T$.  
Local energy (\ref{eq:ham}) is a product 
of $\vS$$\cdot$$\vS '$ and $\Gamma$. 
Thus lowering energy requires either 
$\vS$$\cdot$$\vS '$$\sim$$+1$ and $\Gamma$$\sim$$\Gamma_-$, 
or 
$\vS$$\cdot$$\vS '$$\sim$$\, -1$ and $\Gamma$$\sim$$\Gamma_+$, 
and the latter is more effective since $|\Gamma_-|$$<$$\Gamma_+$.  
Therefore, many bonds realize 
$\Gamma (\vR , \PLUS{\vR}{\va _j})$$\sim$$\Gamma_+$ 
and this requires 
$\vtau (\vR)$$\approx$$\vd _j$ and 
$\vtau (\PLUS{\vR}{\va _j})$$\approx$$\vd _{j+1}$.  
This constrains $\Gamma$ on the other bonds connected 
to the site $\vR$ or $\PLUS{\vR}{\va _j}$.  
This is because 
$\vtau (\vR)$$\cdot$$\vd _{j\pm1}$$\approx$$-$$1/2$ 
and this enforces 
$\Gamma (\vR , \PLUS{\vR}{\va _{j\pm 1}})$$\sim$$($$\gamma$$-$1/2$)^2$, 
\textit{i.e.} 0 for $S$=$1/2$.  
This explains the configuration of isolated nematicity dimers 
in the $S$=$1/2$ case.\cite{Ferrero}
The same argument predicts for our $S$=$3/2$ case that nematicity dimers 
are weakly connected by all the other bonds with $\Gamma$$\sim$$1/16$, 
but the observed snapshot exhibits different correlations. 
This is clearly seen in the distribution function Fig.~\ref{fig4}(b) 
of angle pairs $P(\theta_0,\theta_1)$, where 
$\theta_0$ and $\theta_1$ are 
the angle of $\vtau (\vR)$ and $\vtau (\PLUS{\vR}{\va _1})$, 
respectively. 
The above argument predicts a sharp peak 
at 
$(\theta_0,\theta_1)$=$(\omega_1 ,\omega_2 )$=$(-\pi/3,\pi)$. 
However, the calculated results show that 
the peak splits with a finite deviation.  
We will continue analysis to clarify this point.  

\begin{figure}[bt]
\begin{center}
\includegraphics[width=8.0cm,bb=0 0 1753 839]{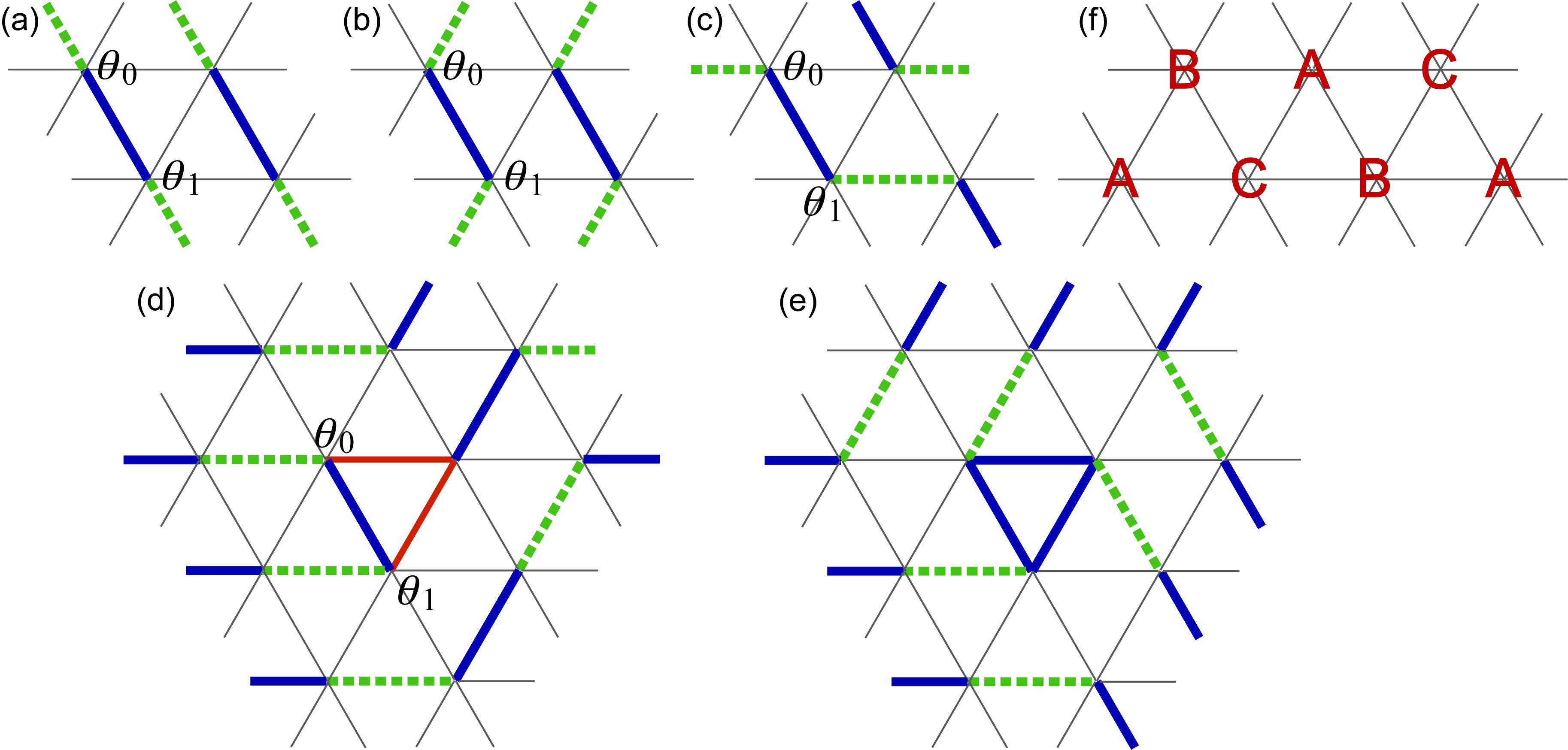}
\end{center}
\caption{(Color online)
Nematicity correlations in some degenerate ground states: 
$(\delta \theta_0, \delta \theta_1)$=
(a)($+$$\Delta$, $-$$\Delta$), 
(b)($+$$\Delta$, $+$$\Delta$), 
(c)($-$$\Delta$, $-$$\Delta$), 
(d)($-$$\Delta$, $+$$\Delta$). 
(e) Mixed with a trimer.  
(f) Three-sublattice structure. 
In (a)-(e), 
strong bonds with $\Gamma_{s}$ are shown 
by blue solid line, while weak links with $\Gamma_{w}$
by green dotted line.  
While the configurations (a)-(c) are periodic, 
(d) is a local realization including 
bonds with $\Gamma$=$-(3/4)^2$ shown in red.  
}
\label{fig5}
\end{figure}

To obtain the lowest-energy configuration of 
spins and nematicities for the $S$=$3/2$ case, 
we have performed numerical annealing using 
a rhombus part with $N$=6$\times$6.  
The initial configuration is the one  
where nematicity dimers are periodically aligned 
with $(\theta_0^{(0)},\theta_1^{(0)})$=$(-\pi/3,\pi)$ 
and a random configuration is set for spins.  
The annealed result is again a periodic pattern 
with 
$(\theta_0^*,\theta_1^*)$=$(\theta_0^{(0)},\theta_1^{(0)})$%
+$(\delta \theta_0,\delta \theta_1)$
shown in Fig.~\ref{fig5}(a), 
but the bonds shown by thin line have $\Gamma$=0 
within numerical accuracy. 
This means that 
$\delta \theta_0$=$-$$\delta \theta_1$=$\Delta$
with 
$\Delta$=$\cos^{-1}\!\gamma$$-$$\pi/3$$\approx$0.0862$\pi$, 
and this leads to 
$\Gamma_{s}$=
$(3\phi /4)^2$$\approx$1.4726
for dimer and 
$\Gamma_{w}$=
$(3/4\phi )^2$$\approx$0.2149 
for weak link, 
where $\phi$$\equiv$$(\mbox{$\sqrt{5}$$+$1})/2$ is the 
golden ratio.  
Spin configuration is N\'eel like along each chain of 
coupled dimers, but has no correlation between 
different chains.  
The corresponding energy 
is $E_0^*$=$-27/32$=$-0.84375$ per site.  
The lowest-energy configuration 
is degenerate, even within periodic ones.  
Figure \ref{fig5} (b) and (c) show some of them with a doubled unit cell, 
and the weak links are now along $\va _2$ and $\va _3$.  

These straight chains of weakly connected dimers do not exhaust 
the degeneracy of the lowest-energy configuration. 
First, these chains can bend with no energy cost.  
Another variant is the one shown in Fig.~\ref{fig5}(d) 
in which weak links connect differently with $\Gamma $$<$0.  
Secondly, one can mix a trimer and it changes the direction 
of dimers as shown in Fig.~\ref{fig5} (e).  
Thirdly, there also exist two-dimensional configurations, 
and a typical example is a 120$^\circ$ state shown in Fig.~\ref{fig5}(f).  
This is a state with three sublattices A-C, 
where cluster spins and nematicities 
exhibit an ordinary 120$^\circ$ order with 
$\theta_A$$-$$\theta_B$$\equiv$$\theta_B$$-$$\theta_C$$\equiv$$2\pi/3$ 
modulo $2\pi$, 
and this state has a continuous degeneracy 
in both spaces.  

Thus, the lowest-energy configuration has very large 
degrees of freedom, supposedly a macroscopic entropy 
as for the $S$=1/2 case,\cite{Mila98} 
but we leave its estimate for a future study.  
The snapshot in Fig.~\ref{fig4}(a) is one example.  
We plot in Fig.~\ref{fig4}(c) the integrated distribution 
of bond and site energies in this snapshot.  
Site energy is defined as half of the sum of the related 6 bond 
energies.  
Bond energies have a large variance and range from 
$-1.5601$ to $0.3883$. 
Many of them are around zero, which means energetically 
disconnected bonds, i.e., missing bonds in the defective network.  
In contract, site energies have only a small variance, 
and its maximum and minimum differ by only 0.0125.  
Its average value is $\bar{E}_{\rm site}$=$-0.84374$, 
almost identical to $E_0^*$ 
considering the effect of nonzero temperature.

\begin{figure}[bt]
\begin{center}
\includegraphics[width=8.5cm,bb=0 0 1364 490]{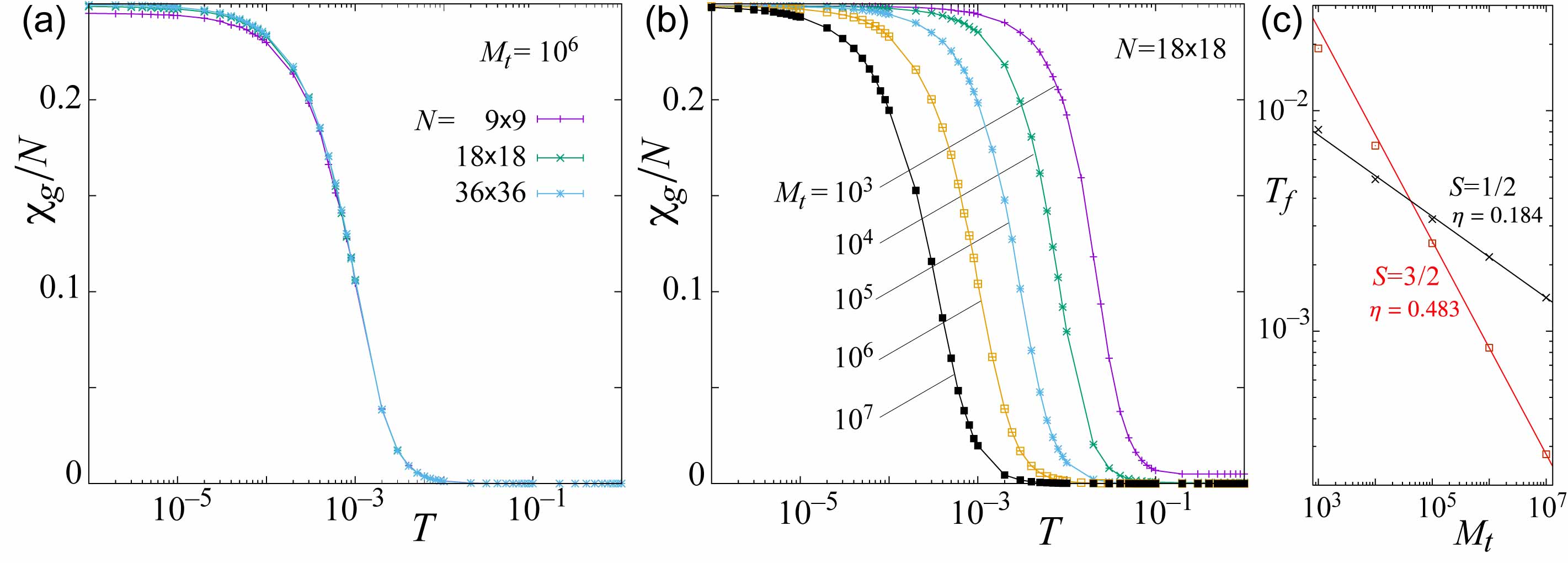}
\end{center}
\caption{(Color online)
Semi-log plot of glass susceptibility of nematicities. 
(a) System size dependence.  
(b) Dependence on $M_t$. 
Data for $M_t$=$10^6$ are a different set from that in (a).  
(c) Scaling of the freezing temperature 
$T_f$ defined by the midpoint of 
each slope in (b).  
}
\label{fig6}
\end{figure}

Now, let us come back to the discussion of  
the low-$T$ value of $C$.  
Since $\vS$ and $\vtau$ are both classical unit vectors, 
Dulong-Petit law predicts that 
each of their angles contributes 1/2 to $C$, 
and this explains the value $\sim$1.5 in Fig.~\ref{fig2}(b). 
By contrast, for $S$=1/2,
$C$=3/2 was found in Ref.~8 
using 3-component $\vtau$,
and our calculation using 2-component $\vtau$ found $C$=1. 
In both cases, 1/2 is missing 
and this is because 
half the spin degrees of freedom are zero modes 
in dimers and trimers.\cite{Ferrero}
The value in Fig.~\ref{fig2}(b) is slightly smaller than 3/2, 
and this reduction is also due to spin zero modes 
in disconnected blobs mixed in the defective network.  
Precise counting of zero modes requires analysis of 
density and shapes of isolated blobs together with 
analysis of local spin structure in each blob.  
This is another challenge to be studied in future.  

Lastly, let us discuss the glassy behavior of nematicities,  
which was reported for the $S$=1/2 case,\cite{Ferrero} 
while spin glass behaviors in undistorted kagome has been 
discussed much longer.\cite{Chandra}
At very low $T$, a crossover takes place into a 
state where nematicities are frozen and 
only spin degrees of freedom fluctuate subject to 
the frozen pattern of effective exchange 
couplings.\cite{Ferrero,Mila02,Cepas}  
This crossover is signaled by the nematicity glass susceptibility 
\begin{equation}
 \chi_g \equiv 
\frac{1}{N} \sum_{\vR , \vR '} 
\left[ \frac{1}{M_t} 
\sum_{t=1}^{M_t} \vtau _t (\vR ) \cdot \vtau _t (\vR ')
\right]^2
\end{equation}
where $\vtau _t (\vR )$ is the variable at Monte-Carlo step $t$.  
Note that $\chi_g$ takes a value of O$(N)$ in states 
where nematicities do not fluctuate with ``time'' $t$, 
which include both regularly ordered and glassy states.  
Since the possibility of regularly ordered state is 
already excluded by the absence of sharp peak in 
$C_{\tau} (\vk )$, a nonvanishing value of $\chi_g /N$ 
means a glassy behavior of nematicity.  
Figure \ref{fig6} shows its temperature dependence 
for various values of system size and $M_t$.  
Panel (a) shows that 
$\chi_g /N$ grows noticeably at $T$$\lesssim$$10^{-3}$, 
and this is an evidence of \textit{freezing} to a glassy state.  
The system size dependence is very small except in the region 
of lowest temperatures, and the value increases with $N$ there. 
This indicates that this glassy behavior survives in the $N$=$\infty$ 
limit, but it is premature to conclude this.  

Figure \ref{fig6}(b) shows the results with varying 
measurement time $M_t$.  
With increasing $M_t$, freezing starts 
at a lower temperature, 
and this agrees with the previous study for the $S$=1/2 case.\cite{Ferrero}
Each time multiplying $M_t$ by the factor 10, the whole curve 
moves with a nearly constant shift 
towards the lower-temperature side in this semi-log plot. 
This implies the characteristic freezing temperature scales 
as $T_f$$\sim$$M_t^{-\eta}$, and the exponent 
is estimated in Fig.~\ref{fig6}(c) as $\eta$$\approx$0.483.  
Therefore, the defective network of nematicity 
has slow Monte Carlo dynamics, and its time 
scale $t_f$ grows with lowering temperature as 
$t_f$$\sim$$T^{-1/\eta}$.
We have also calculated the results for the $S$=1/2 case.  
The difference from the $S$=3/2 case is only quantitative 
and the exponent is smaller $\eta$$\approx$0.184, while 
the temperature dependence was claimed activation type 
in the previous study.\cite{Ferrero}
This nonuniversal exponent may imply that it is only a transient behavior 
but this scaling holds well down to the lowest temperatures 
in our simulation.

In this letter, we have theoretically studied 
the trimerized kagome antiferromagnetic Heisenberg model 
with half integer spin $S$ 
in the region of large trimerization.  
For general $S$, 
we constructed its low-energy effective Hamiltonian 
in terms of spin and nematicity of each triangular unit.  
We then performed its classical Monte Carlo simulations 
for the $S$=3/2 case and studied its low temperature properties. 
We found that nematicities form local dimers and trimers 
as for the $S$=1/2 case, but they are weakly coupled to 
form a defective planar network mixed with disconnected 
small blobs.  
Nematicities also show glassy slow Monte Carlo dynamics, and 
its characteristic time scale grows with temperature 
following a power law.  
Concerning spin interactions, they are modulated in space 
according to the pattern of defective network of 
nearly frozen nematicities, and many bonds 
are effectively broken.  
Spin correlations grow at low temperatures 
but exhibit no periodic pattern detected by a sharp peak 
in $C_S (\vk)$.  
Therefore, the low-energy state is 
a strongly correlated classical spin liquid 
on a defective network of nematicities with glassy slow dynamics. 
Since nematicity represents breaking of the triangular symmetry 
of each unit, it couples to local lattice distortion. 
It is worthwhile to note that 
one can detect these nematicities in principle  
by measuring E-mode of local lattice distortion.  

\begin{acknowledgment}
The major part of the numerical calculations were performed 
at the Supercomputer Center, ISSP, at the University of Tokyo.  
\end{acknowledgment}


\begin{thebibliography}{99}
\bibitem{Mambrini}
Reference 8 remarked that 
the same type of effective Hamiltonian was 
derived for general half integer $S$ by M.~Mambrini, 
but we have found no related publication.
\end{thebibliography}
\end{document}